\documentclass[12pt]{article}
\usepackage{epsfig}
\usepackage{amsmath}
\usepackage{hhline}
\usepackage{amssymb}
\usepackage{times}
\usepackage{cite}
\usepackage{rotating}
\newcommand{\rb}[1]{\raisebox{1.5ex}[-1.5ex]{#1}}
\newcommand{\F}{$ F_{2}(x,Q^2)\:$} 

\newcommand{\Fc}{$ F_{2}\,$}

\newcommand{\gv}{GeV$^2\,$}

\newcommand{\FLc}{$ F_{L}\,$}

\newcommand{\beq}{\begin{equation}}                       
\newcommand{\eeq}{\end{equation}}                       
\newcommand{\PO}{I\!\!P }

\newcommand{\llam}{$\lambda(x,Q^2)\:$}
\newcommand{\lam}{$\lambda(Q^2)\:$}
\newlength{\dinwidth}                                                          
\newlength{\dinmargin}                                                         
\begin{document}              
\noindent                                                                 
\begin{titlepage}                                                    
\begin{flushleft}
DESY-01-104  \hfill  ISSN 0418-9833 \\
July 2001
\end{flushleft}

\vspace*{3.cm} 

\begin{center}                                                                 

\begin{Large}                                                                  
{\bf On the Rise 
of  the \\  Proton  Structure Function 
{\boldmath \Fc} \\  Towards Low {\boldmath $x$}  \\}
\vspace*{1cm}                                                                 
H1 Collaboration
\\                                                          
\end{Large}
\vspace*{2.cm} {\bf Abstract:   }
\begin{quotation}               
\noindent
A measurement of the derivative $(\partial \ln F_2 / \partial \ln
x)_{Q^2} \equiv -\lambda(x,Q^2)$ of the proton structure function
$F_2$ is presented in the low $x$ domain of deeply inelastic
positron--proton scattering. For $ 5\cdot 10^{-5} \leq x \leq 0.01$
and $Q^2\geq 1.5\,{\rm GeV}^2$, $\lambda(x,Q^2)$ is found to be
independent of $x$ and to increase linearly with $\ln Q^2$.
\end{quotation}
\vspace*{2.cm} 
{\it Submitted to Phys. Lett. B}
\end{center}                                                                
\cleardoublepage                     
\end{titlepage}
\begin{flushleft}

C.~Adloff$^{33}$,              
V.~Andreev$^{24}$,             
B.~Andrieu$^{27}$,             
T.~Anthonis$^{4}$,             
V.~Arkadov$^{35}$,             
A.~Astvatsatourov$^{35}$,      
A.~Babaev$^{23}$,              
J.~B\"ahr$^{35}$,              
P.~Baranov$^{24}$,             
E.~Barrelet$^{28}$
% ...

\bigskip{\it
% ...
$ ^{4}$ Inter-University Institute for High Energies ULB-VUB, Brussels;
          Universitaire Instelling Antwerpen, Wilrijk; Belgium$^{ c}$ \\
% ...
$ ^{23}$ Institute for Theoretical and Experimental Physics,
          Moscow, Russia$^{ l}$ \\
 $ ^{24}$ Lebedev Physical Institute, Moscow, Russia$^{ e,h}$ \\
% ...
$ ^{27}$ LPNHE, Ecole Polytechnique, IN2P3-CNRS, Palaiseau, France \\
 $ ^{28}$ LPNHE, Universit\'{e}s Paris VI and VII, IN2P3-CNRS,
          Paris, France \\
% ...
$ ^{33}$ Fachbereich Physik, Bergische Universit\"at Gesamthochschule
          Wuppertal, Wuppertal, Germany \\
% ...
$ ^{35}$ DESY, Zeuthen, Germany \\
% ...
$ ^c$ Supported by FNRS-NFWO, IISN-IIKW \\
% ...
$ ^e$ Supported by the Deutsche Forschungsgemeinschaft \\
% ...
$ ^h$ Supported by Russian Foundation for Basic Research
      grant no. 96-02-00019 \\
% ...
$ ^l$ Partially Supported by Russian Foundation
      for Basic Research, grant    no. 00-15-96584 \\
}

\end{flushleft}
\newpage
%
%
%
\noindent
The inclusive cross section for deeply inelastic lepton-proton
scattering is governed by the proton structure function \F. Because of
the large centre-of-mass energy squared, $s \simeq 10^5 $~GeV$^2$, the
$ep$ collider HERA has accessed the region of low Bjorken $x$, $x >
Q^2/s > 10^{-5}$, for four-momentum transfers squared $Q^2 > 1 $
GeV$^2$. One of the first observations at HERA was of a substantial
rise of \Fc with decreasing $x$~\cite{f292}. However, this rise may be
limited at very low $x$ by unitarity constraints.

Perturbative Quantum Chromodynamics (QCD) provides a rigorous and
successful theoretical description of the $Q^2$ dependence of \F in
deeply inelastic scattering.  In the double asymptotic limit, the
DGLAP evolution equations~\cite{dglap} can be solved~\cite{das} and
\Fc\ is expected to rise approximately as a power of $x$ towards low
$x$. A power behaviour is also predicted in BFKL theory~\cite{bfkl}.
The rise is expected eventually to be limited by gluon self interactions
in the nucleon~\cite{glr}.

Recently the H1 Collaboration has presented~\cite{paper} a new
measurement of \F in the kinematic range $3 \cdot 10^{-5} \le x \le
0.2$ and $1.5 \le Q^2 \le 150 $ GeV$^2$ based on data taken in the
years 1996/97 with a positron beam energy $E_e = 27.6$ GeV and a
proton beam energy $E_p = 820$ GeV.  The high accuracy of these data
allows the derivative
\begin{equation}
    \left( \frac{\partial \ln F_2(x,Q^2)}{\partial \ln x}\right)_{Q^2}  \equiv 
    - \lambda(x,Q^2)
\end{equation} 
to be measured as a function both of $Q^2$ and of $x$ for the first
time.  Use of this quantity for investigating the behaviour of \Fc at
low $x$ was suggested in~\cite{Navel}.

Here results are presented of a measurement of this derivative in the
full kinematic range available. Data points at adjacent values of $x$
and at fixed $Q^2$ are used~\cite{paper} taking account of the full
error correlations and the spacing between the $x$ values.  The
results\footnote{Note that derivatives at adjacent $x$ values are thus
anti-correlated. 
The data points at $Q^2=150$~\gv are
  obtained from the H1 measurement~\cite{fbea}.}  obtained are
presented in Table~\ref{tabder}.  The sensitivity of the derivative to
the uncertainty of the structure function \FLc~\cite{paper} throughout
the measured kinematic range is estimated to be much smaller than the
total systematic error at the lowest values of $x$ and is negligible
elsewhere.

\begin{figure}[ht]
\epsfig{file=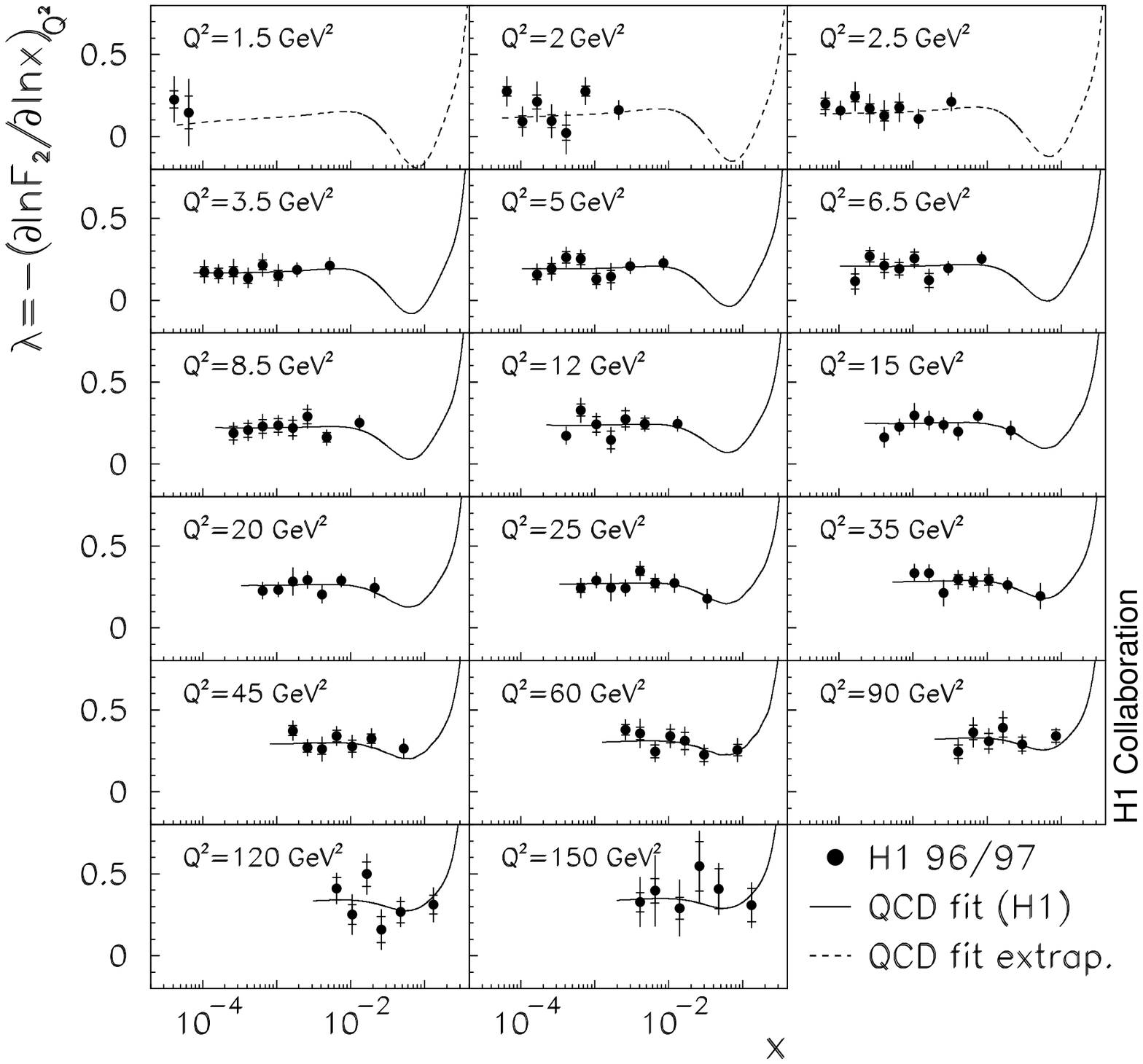,width=15.cm}
  \caption{ Measurement of the function \llam: the inner error bars 
    represent the statistical uncertainty; the full
    error bars include the systematic uncertainty added in quadrature;
    the solid curves represent the NLO QCD fit to the H1 cross section
    data described in~\cite{paper}; the dashed curves represent the
    extrapolation of the QCD fit below $Q^2 = 3.5 $ GeV$^2$.}
  \protect\label{lamx}
\end{figure}

As can be seen in Figure~\ref{lamx}, the derivative \llam is
independent of $x$ for $x \lesssim 0.01$ to within the experimental
accuracy.  This implies that the $x$ dependence of \Fc at low $x$ is
consistent with a power law, $F_2 \propto x^{-\lambda}$, for fixed
$Q^2$, and that the rise of \Fc, i.e.  $(\partial F_2 / \partial
x)_{Q^2}$, is proportional to $F_2/x$.  There is no experimental
evidence that this behaviour changes in the measured kinematic range.

The derivative is well described by the NLO QCD fit to the H1
cross-section data~\cite{paper}, see Figure~\ref{lamx}. In DGLAP QCD,
for $Q^2 > 3$ GeV$^2$, the low $x$ behaviour is driven solely by the
gluon field, since quark contributions to the scaling violations of
\Fc\ are negligible.  At larger $x$ the transition to the
valence-quark region causes a strong dependence of $\lambda$ on $x$ as
indicated by the QCD curves in Figure~\ref{lamx}.

\begin{figure}[ht]
\epsfig{file=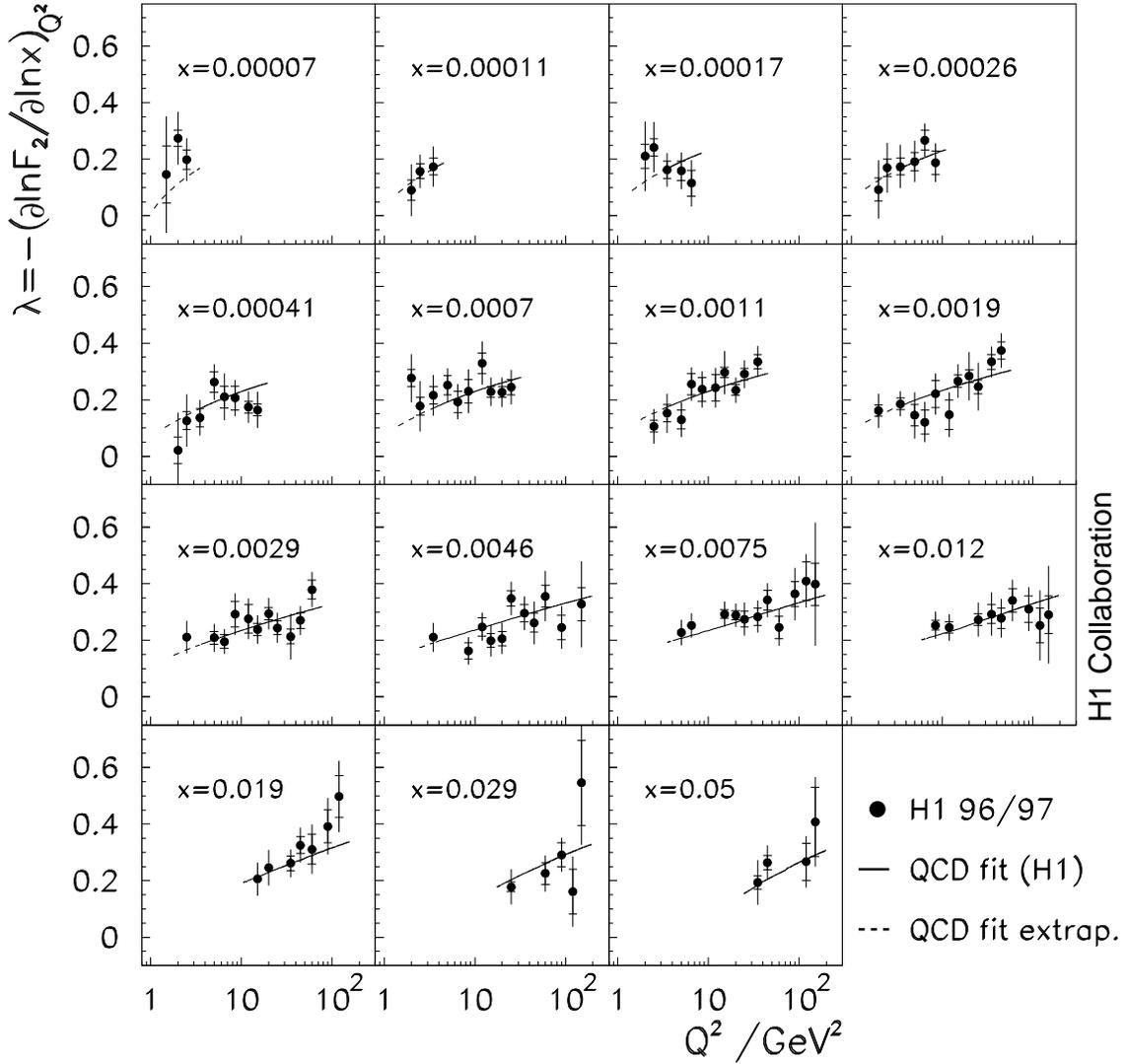,width=15.cm}
  \caption{ Measurement of the function \llam: the inner error bars
    represent the statistical uncertainty; the
    full error bars include  the systematic uncertainty added in
    quadrature; the solid curves represent the NLO QCD fit to the H1
    cross section data described in~\cite{paper}; the minimum $Q^2$
    value of the data included in this fit is $Q^2 = 3.5 $
    GeV$^2$.  } \protect\label{lamq2}
\end{figure}

Figure~\ref{lamq2} shows the measured derivative as a function of
$Q^2$ for different $x$ values.  The derivative is observed to rise
approximately logarithmically with $Q^2$. It can be represented by a
function \lam which is independent of $x$ within the experimental
accuracy.

\begin{figure}[ht]
 \epsfig{file=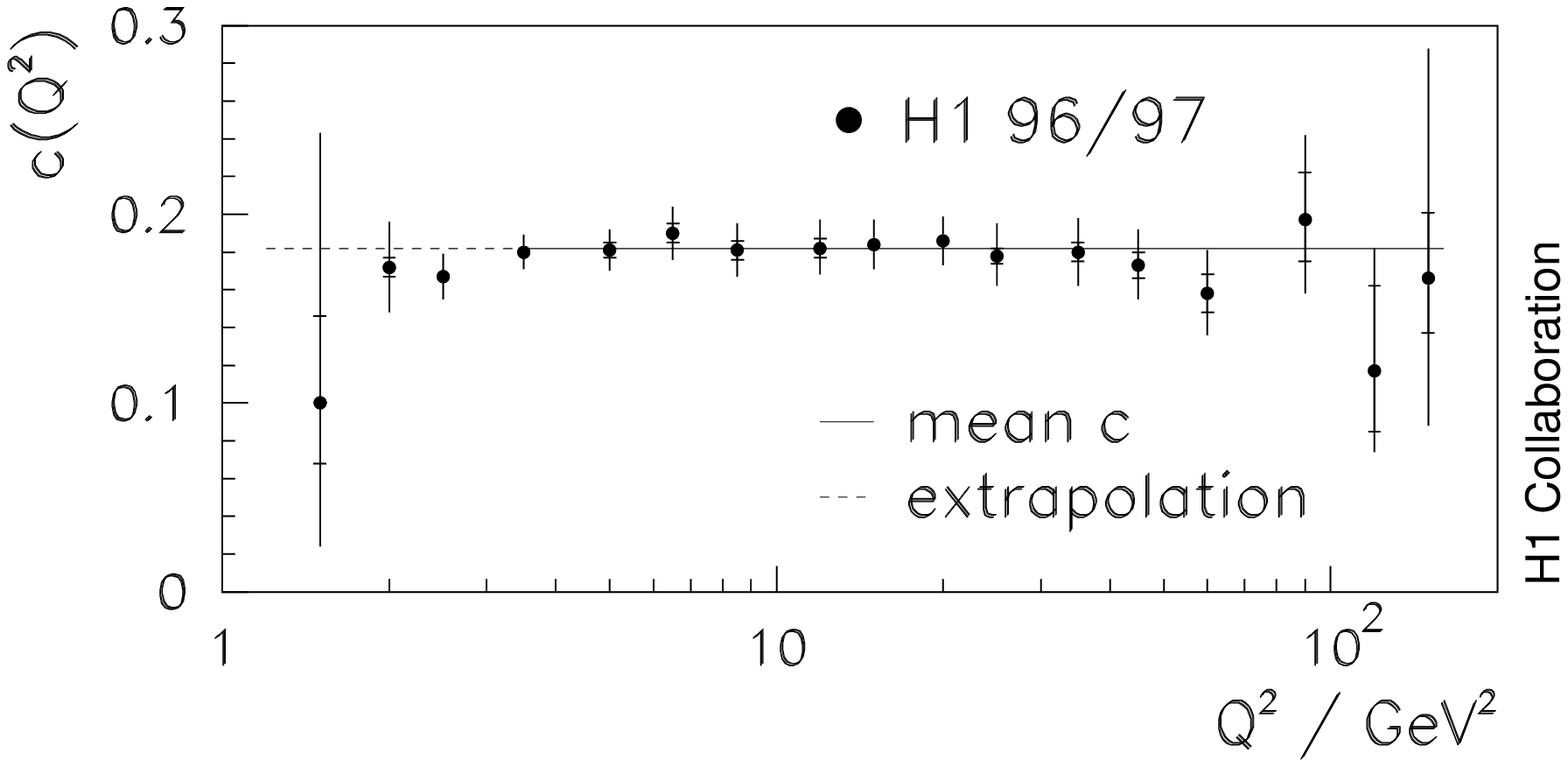,width=13.7cm}
 \epsfig{file=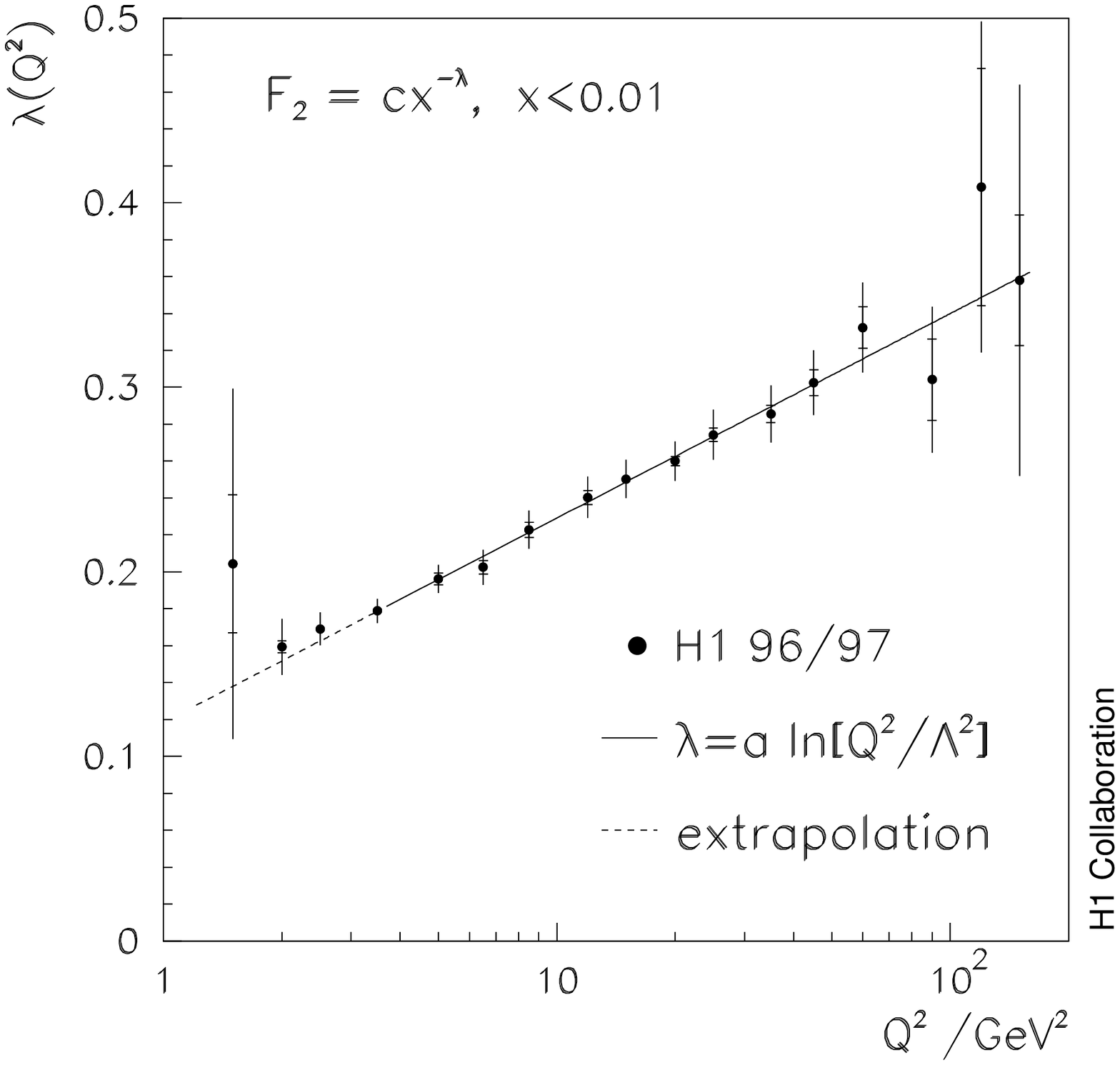,width=13.7cm}
  \caption{ Determination of  the coefficients $c(Q^2)$ (upper plot) and of the 
    exponents \lam (lower plot) from fits of the form $F_2(x,Q^2) =
    c(Q^2) x^{-\lambda(Q^2) }$ to the H1 structure function
    data~\cite{paper} for $x \leq 0.01$; the inner error bars
    illustrate the statistical uncertainties, the full error bars
    represent the statistical and systematic uncertainties added
    in quadrature. The straight lines represent the mean coefficient
    $c$ (upper plot) and a fit of the form $ a \ln
    [Q^2/\Lambda^2]$ (lower plot), respectively, using data for $Q^2
    \geq 3.5$ GeV$^2$.}
  \protect\label{lammax}
\end{figure}

The function \lam is determined from fits of the form $F_2(x,Q^2) =
c(Q^2) x^{-\lambda(Q^2)}$ to the H1 structure function data,
restricted to the region $x \leq 0.01$.  The results for $c(Q^2)$ and
\lam are presented in Table~\ref{tablam} and shown in
Figure~\ref{lammax}.  The coefficients $c(Q^2)$ are approximately
independent of $Q^2$ with a mean value of 0.18.  As can be seen,
$\lambda(Q^2)$ rises approximately linearly with $\ln Q^2$. This
dependence can be represented as $\lambda(Q^2) = a \cdot \ln
[Q^2/\Lambda^2]$, see Figure~\ref{lammax}.  The coefficients are
$a=0.0481 \pm 0.0013(\rm stat)\pm 0.0037(\rm syst)$ and $\Lambda = 292
\pm 20(\rm stat) \pm 51(\rm syst)$~MeV, obtained for $Q^2 \geq 3.5$
GeV$^2$. The values of \lam are more accurate than data hitherto
published by the H1~\cite{h1svx} and ZEUS~\cite{zeuslam}
Collaborations.

Below the deeply inelastic region, for fixed $Q^2 < 1 $~GeV$^2$, the
simplest Regge phenomenology predicts that \F $\propto x^{-\lambda}$
where $\lambda = \alpha_{\PO}(0)-1 \simeq 0.08$ is given by the
Pomeron intercept independently of $x$ and $Q^2$~\cite{dl}.  When
extrapolating the function \lam into the lower $Q^2$ region it has the
value of $0.08$ at $Q^2 = 0.45$~GeV$^2$, see also \cite{zeuslam}.

To summarise, the derivative $(\partial \ln F_2 / \partial \ln
x)_{Q^2}$ is measured as a function both of $x$ and of $Q^2$ and is
observed to be independent of Bjorken $x$ for $x \lesssim 0.01$ and
$Q^2$ between 1.5 and 150~\gv. Thus the behaviour of \Fc at low $x$ is
consistent with a dependence $F_2(x,Q^2)~=~c(Q^2)~x^{-\lambda(Q^2)}$
throughout that region.  At low $x$, the
exponent $\lambda$ is observed to rise linearly with $\ln Q^2$ and the
coefficient $c$ is independent of $Q^2$ to within the experimental
accuracy.  There is no sign that this behaviour changes within the
kinematic range of deeply inelastic scattering explored.

%
{\bf Acknowledgements}
\normalsize
\noindent 
We are very grateful to the HERA machine group whose
outstanding efforts made this experiment possible. We acknowledge the
support of the DESY technical staff. We appreciate the substantial
effort of the engineers and technicians who constructed and maintain
the detector. We thank the funding agencies for financial support of
this experiment.  We wish to thank the DESY directorate for the
support and hospitality extended to the non-DESY members of the
collaboration. 
%

%
%
%
%
%
%
%
%
\begin{table}[h]  
\begin{tabular}{cc}
\begin{scriptsize}
  \begin{tabular}[t]{|c|l|c|r|r|r|}
  \hline
     \multicolumn{1}{|c|}{$Q^2$}   & & & & & \\
   $[$GeV$^2]$ & \multicolumn{1}{|c|}{\rb{$x$}} &  {\rb{$\lambda$}} & 
    \rb{$\delta_{sta}$} 
  & \rb{$\delta_{sys}$} & \rb{$\delta_{tot}$}    \\
  \hline
      1.5 & 0.000041 & 0.225 & 0.052 & 0.131 & 0.141 \\
    1.5 & 0.000065 & 0.146 & 0.101 & 0.180 & 0.206 \\  
    2.0 & 0.000065 & 0.274 & 0.029 & 0.089 & 0.093 \\  
    2.0 & 0.000105 & 0.090 & 0.036 & 0.085 & 0.092 \\  
    2.0 & 0.000165 & 0.211 & 0.043 & 0.116 & 0.124 \\  
    2.0 & 0.000260 & 0.093 & 0.040 & 0.096 & 0.104 \\  
    2.0 & 0.000410 & 0.022 & 0.047 & 0.124 & 0.132 \\  
    2.0 & 0.000750 & 0.276 & 0.030 & 0.080 & 0.086 \\  
    2.0 & 0.00210  & 0.161 & 0.020 & 0.057 & 0.060 \\  
    2.5 & 0.000065 & 0.199 & 0.034 & 0.068 & 0.076 \\  
    2.5 & 0.000105 & 0.158 & 0.026 & 0.053 & 0.059 \\  
    2.5 & 0.000165 & 0.242 & 0.031 & 0.084 & 0.089 \\  
    2.5 & 0.000260 & 0.170 & 0.030 & 0.082 & 0.088 \\  
    2.5 & 0.000410 & 0.126 & 0.032 & 0.088 & 0.093 \\  
    2.5 & 0.000650 & 0.177 & 0.032 & 0.084 & 0.090 \\  
    2.5 & 0.00119  & 0.106 & 0.021 & 0.058 & 0.061 \\  
    2.5 & 0.00329  & 0.211 & 0.013 & 0.057 & 0.058 \\  
    3.5 & 0.000105 & 0.174 & 0.030 & 0.065 & 0.071 \\  
    3.5 & 0.000165 & 0.163 & 0.031 & 0.050 & 0.059 \\  
    3.5 & 0.000260 & 0.174 & 0.030 & 0.071 & 0.077 \\  
    3.5 & 0.000410 & 0.136 & 0.032 & 0.054 & 0.063 \\  
    3.5 & 0.000650 & 0.216 & 0.031 & 0.065 & 0.072 \\  
    3.5 & 0.00105  & 0.152 & 0.031 & 0.063 & 0.070 \\  
    3.5 & 0.00190  & 0.186 & 0.021 & 0.039 & 0.044 \\  
    3.5 & 0.00525  & 0.210 & 0.012 & 0.050 & 0.051 \\  
    5.0 & 0.000165 & 0.159 & 0.035 & 0.055 & 0.065 \\  
    5.0 & 0.000260 & 0.192 & 0.033 & 0.065 & 0.073 \\  
    5.0 & 0.000410 & 0.262 & 0.036 & 0.052 & 0.063 \\  
    5.0 & 0.000650 & 0.251 & 0.034 & 0.049 & 0.060 \\  
    5.0 & 0.00105  & 0.129 & 0.033 & 0.051 & 0.061 \\  
    5.0 & 0.00165  & 0.145 & 0.038 & 0.072 & 0.081 \\  
    5.0 & 0.00299  & 0.209 & 0.022 & 0.044 & 0.049 \\  
    5.0 & 0.00849  & 0.227 & 0.013 & 0.042 & 0.044 \\  
    6.5 & 0.000165 & 0.115 & 0.046 & 0.068 & 0.082 \\  
    6.5 & 0.000260 & 0.268 & 0.036 & 0.047 & 0.059 \\  
    6.5 & 0.000410 & 0.210 & 0.039 & 0.073 & 0.082 \\  
    6.5 & 0.000650 & 0.193 & 0.038 & 0.049 & 0.062 \\  
    6.5 & 0.00105  & 0.255 & 0.037 & 0.047 & 0.060 \\  
    6.5 & 0.00165  & 0.121 & 0.042 & 0.058 & 0.072 \\  
    6.5 & 0.00299  & 0.195 & 0.024 & 0.037 & 0.044 \\  
    6.5 & 0.00849  & 0.253 & 0.014 & 0.040 & 0.043 \\  
    8.5 & 0.000260 & 0.188 & 0.042 & 0.054 & 0.068 \\  
    8.5 & 0.000410 & 0.206 & 0.042 & 0.048 & 0.064 \\  
    8.5 & 0.000650 & 0.230 & 0.042 & 0.064 & 0.077 \\  
    8.5 & 0.00105  & 0.237 & 0.042 & 0.049 & 0.064 \\  
    8.5 & 0.00165  & 0.220 & 0.047 & 0.054 & 0.072 \\  
    8.5 & 0.00260  & 0.292 & 0.045 & 0.058 & 0.073 \\  
    8.5 & 0.00475  & 0.162 & 0.028 & 0.040 & 0.049 \\  
    8.5 & 0.0132   & 0.253 & 0.017 & 0.045 & 0.048 \\  
   12.0 & 0.000410 & 0.174 & 0.020 & 0.052 & 0.056 \\  
   12.0 & 0.000650 & 0.330 & 0.036 & 0.067 & 0.076 \\  
   12.0 & 0.00105  & 0.243 & 0.047 & 0.052 & 0.070 \\  
   12.0 & 0.00165  & 0.147 & 0.053 & 0.059 & 0.079 \\  
   12.0 & 0.00260  & 0.276 & 0.050 & 0.053 & 0.073 \\  
   12.0 & 0.00475  & 0.247 & 0.032 & 0.038 & 0.049 \\  
   12.0 & 0.0132   & 0.245 & 0.019 & 0.043 & 0.047 \\  
   15.0 & 0.000410 & 0.164 & 0.020 & 0.060 & 0.064 \\  
   15.0 & 0.000650 & 0.227 & 0.018 & 0.047 & 0.050 \\ 
   15.0 & 0.00105  & 0.296 & 0.018 & 0.074 & 0.076 \\   
   15.0 & 0.00165  & 0.266 & 0.021 & 0.055 & 0.059 \\   

  \hline
  \end{tabular}
  \begin{tabular}[t]{|c|l|c|r|r|r|}
  \hline 
    \multicolumn{1}{|c|}{$Q^2$}   & & & & & \\
   $[$GeV$^2]$ & \multicolumn{1}{|c|}{\rb{$x$}} &  {\rb{$\lambda$}} & 
\rb{$\delta_{sta}$} 
  & \rb{$\delta_{sys}$} & \rb{$\delta_{tot}$}    \\
  \hline
     15.0 & 0.00260  & 0.238 & 0.021 & 0.045 & 0.050 \\  
   15.0 & 0.00410  & 0.199 & 0.024 & 0.050 & 0.056 \\  
   15.0 & 0.00750  & 0.292 & 0.016 & 0.041 & 0.044 \\  
   15.0 & 0.0210   & 0.206 & 0.013 & 0.057 & 0.059 \\  
   20.0 & 0.000650 & 0.227 & 0.019 & 0.048 & 0.052 \\  
   20.0 & 0.00105  & 0.234 & 0.019 & 0.041 & 0.045 \\  
   20.0 & 0.00165  & 0.284 & 0.022 & 0.082 & 0.085 \\  
   20.0 & 0.00260  & 0.293 & 0.022 & 0.051 & 0.055 \\  
   20.0 & 0.00410  & 0.206 & 0.025 & 0.049 & 0.055 \\  
   20.0 & 0.00750  & 0.289 & 0.016 & 0.038 & 0.041 \\  
   20.0 & 0.0210   & 0.246 & 0.012 & 0.063 & 0.064 \\  
   25.0 & 0.000650 & 0.244 & 0.027 & 0.055 & 0.061 \\  
   25.0 & 0.00105  & 0.291 & 0.020 & 0.045 & 0.050 \\  
   25.0 & 0.00165  & 0.246 & 0.024 & 0.081 & 0.084 \\  
   25.0 & 0.00260  & 0.244 & 0.024 & 0.047 & 0.053 \\  
   25.0 & 0.00410  & 0.348 & 0.027 & 0.053 & 0.059 \\  
   25.0 & 0.00650  & 0.274 & 0.028 & 0.051 & 0.058 \\  
   25.0 & 0.0119   & 0.273 & 0.020 & 0.056 & 0.060 \\  
   25.0 & 0.0329   & 0.178 & 0.016 & 0.060 & 0.062 \\  
   35.0 & 0.00105  & 0.335 & 0.025 & 0.049 & 0.055 \\  
   35.0 & 0.00165  & 0.334 & 0.026 & 0.047 & 0.054 \\  
   35.0 & 0.00260  & 0.213 & 0.026 & 0.077 & 0.081 \\  
   35.0 & 0.00410  & 0.295 & 0.030 & 0.051 & 0.059 \\  
   35.0 & 0.00650  & 0.283 & 0.030 & 0.048 & 0.057 \\  
   35.0 & 0.0105   & 0.293 & 0.033 & 0.070 & 0.077 \\  
   35.0 & 0.0190   & 0.261 & 0.026 & 0.042 & 0.049 \\  
   35.0 & 0.0525   & 0.194 & 0.023 & 0.076 & 0.079 \\  
   45.0 & 0.00165  & 0.374 & 0.031 & 0.053 & 0.061 \\  
   45.0 & 0.00260  & 0.270 & 0.028 & 0.043 & 0.052 \\  
   45.0 & 0.00410  & 0.262 & 0.033 & 0.068 & 0.075 \\  
   45.0 & 0.00650  & 0.342 & 0.035 & 0.049 & 0.060 \\  
   45.0 & 0.01050  & 0.277 & 0.037 & 0.058 & 0.069 \\  
   45.0 & 0.0190   & 0.325 & 0.029 & 0.055 & 0.062 \\  
   45.0 & 0.0525   & 0.263 & 0.025 & 0.056 & 0.061 \\  
   60.0 & 0.00260  & 0.379 & 0.033 & 0.054 & 0.063 \\  
   60.0 & 0.00410  & 0.356 & 0.038 & 0.081 & 0.089 \\  
   60.0 & 0.00650  & 0.246 & 0.040 & 0.051 & 0.064 \\  
   60.0 & 0.0105   & 0.340 & 0.042 & 0.061 & 0.074 \\  
   60.0 & 0.0165   & 0.311 & 0.053 & 0.070 & 0.088 \\  
   60.0 & 0.0299   & 0.225 & 0.039 & 0.052 & 0.065 \\  
   60.0 & 0.0849   & 0.255 & 0.035 & 0.064 & 0.073 \\  
   90.0 & 0.00410  & 0.245 & 0.043 & 0.063 & 0.076 \\  
   90.0 & 0.00650  & 0.363 & 0.044 & 0.081 & 0.092 \\  
   90.0 & 0.0105   & 0.310 & 0.047 & 0.061 & 0.077 \\  
   90.0 & 0.0165   & 0.392 & 0.058 & 0.081 & 0.100 \\  
   90.0 & 0.0299   & 0.291 & 0.042 & 0.043 & 0.060 \\  
   90.0 & 0.0849   & 0.340 & 0.039 & 0.035 & 0.052 \\  
  120.0 & 0.00650  & 0.410 & 0.068 & 0.066 & 0.095 \\  
  120.0 & 0.0105   & 0.252 & 0.061 & 0.107 & 0.124 \\  
  120.0 & 0.0165   & 0.498 & 0.074 & 0.102 & 0.126 \\  
  120.0 & 0.0260   & 0.161 & 0.079 & 0.095 & 0.124 \\  
  120.0 & 0.0475   & 0.267 & 0.065 & 0.063 & 0.091 \\  
  120.0 & 0.132    & 0.312 & 0.056 & 0.091 & 0.106 \\  
  150.0 & 0.00410  & 0.327 & 0.059 & 0.141 & 0.153 \\
  150.0 & 0.00650  & 0.397 & 0.075 & 0.204 & 0.218 \\
  150.0 & 0.01400  & 0.291 & 0.067 & 0.159 & 0.173 \\
  150.0 & 0.0260   & 0.545 & 0.150 & 0.158 & 0.218 \\  
  150.0 & 0.0475   & 0.408 & 0.122 & 0.102 & 0.159 \\  
  150.0 & 0.132    & 0.308 & 0.101 & 0.096 & 0.139 \\ 

  \hline
  \end{tabular}
\end{scriptsize}
\end{tabular}
\caption{ Measurement of the derivative $\lambda = -(\partial \ln F_2 / 
          \partial \ln x)_{Q^2}$ at fixed $Q^2$.  For the systematic 
uncertainties 
  the correlations between adjacent $x$ values are taken into account.
The total error is the squared sum of the statistical and 
systematic uncertainties, given as absolute values.}
\protect\label{tabder}
\end{table}
\newpage
%
%
\begin{table}[h] \centering 
\begin{tabular}{|c|l|r|r|l|l|l|}
\hline
$Q^2 [GeV^2]$ &  $c$ & $\delta_{sta}^c$ & $\delta_{tot}^c$ & $ \lambda$ & $\delta_{sta}^{\lambda}$ & $\delta_{tot}^{\lambda}$    \\ \hline
\hline
          &            &$+$ 0.05  &$+$ 0.14  &           &           &           \\
 \rb{1.5} & \rb{0.10}  &$-$ 0.03  &$-$ 0.08  &\rb{0.20} &\rb{0.04} &\rb{0.10} \\ \hline
  2.0     &  0.172     & 0.005 & 0.024 & 0.159     &  0.003    &  0.015    \\ \hline
  2.5     &  0.167     & 0.003 & 0.012 & 0.169     &  0.002    &  0.009    \\ \hline
  3.5     &  0.180     & 0.003 & 0.009 & 0.179     &  0.002    &  0.007    \\ \hline
  5.0     &  0.181     & 0.004 & 0.011 & 0.196     &  0.003    &  0.008    \\ \hline
  6.5     &  0.190     & 0.005 & 0.014 & 0.202     &  0.004    &  0.009    \\ \hline
  8.5     &  0.181     & 0.005 & 0.014 & 0.223     &  0.004    &  0.010    \\ \hline
 12.0     &  0.182     & 0.005 & 0.015 & 0.240     &  0.004    &  0.011    \\ \hline
 15.0     &  0.184     & 0.003 & 0.013 & 0.250     &  0.002    &  0.010    \\ \hline
 20.0     &  0.186     & 0.003 & 0.013 & 0.260     &  0.003    &  0.011    \\ \hline
 25.0     &  0.178     & 0.004 & 0.017 & 0.274     &  0.004    &  0.014    \\ \hline
 35.0     &  0.180     & 0.005 & 0.018 & 0.286     &  0.005    &  0.016    \\ \hline
 45.0     &  0.173     & 0.007 & 0.019 & 0.302     &  0.007    &  0.017    \\ \hline
 60.0     &  0.158     & 0.010 & 0.023 & 0.332     &  0.011    &  0.024    \\ \hline
          &            &$+$ 0.025 &$+$ 0.045 &           &           &           \\
\rb{90.0} &\rb{0.197}  &$-$ 0.022 &$-$ 0.039 &\rb{0.304} &\rb{0.022} &\rb{0.040} \\ \hline
          &            &$+$ 0.045 &$+$ 0.065 &           &           &           \\
\rb{120.0}&\rb{0.117}  &$-$ 0.032 &$-$ 0.043 &\rb{0.408} &\rb{0.064} &\rb{0.089} \\ \hline
          &            &$+$ 0.04 &$+$ 0.12  &            &           &           \\
\rb{150.0}&\rb{0.17}   &$-$ 0.03 &$-$ 0.08  &\rb{0.36}   &\rb{0.04}  &\rb{0.11} \\

\hline
\end{tabular}
\caption{The coefficients $c$ and
 exponents $\lambda$ from fits of the 
form $F_2(x,Q^2) = c(Q^2)x^{-\lambda(Q^2)}$
 using  H1 \Fc 
data~\cite{paper}, for $x \leq 0.01$, taking into account the 
systematic error correlations. Here $\delta_{sta}$
denotes the statistical uncertainty and $\delta_{tot}$
comprises all uncertainties added in quadrature. 
The uncertainties are given as absolute values.
They are symmetric to very good approximation, 
apart from the uncertainties of the coefficient $c(Q^2)$
at the edges of the $Q^2$ region.
} \label{tablam}
\end{table}
\newpage

\end{document}